\documentclass[10pt,english,aps,manuscript,preprint, prl,floatfix,showpacs,amsmath,amssymb,reprint]{revtex4-1}
\usepackage[T1]{fontenc}
\usepackage[latin9]{inputenc}
\usepackage{color}
\usepackage{graphicx}
\usepackage{esint}
\usepackage[normalem]{ulem}
\makeatletter

\@ifundefined{textcolor}{}
{%
 \definecolor{BLACK}{gray}{0}
 \definecolor{WHITE}{gray}{1}
 \definecolor{RED}{rgb}{1,0,0}
 \definecolor{GREEN}{rgb}{0,1,0}
 \definecolor{BLUE}{rgb}{0,0,1}
 \definecolor{CYAN}{cmyk}{1,0,0,0}
 \definecolor{MAGENTA}{cmyk}{0,1,0,0}
 \definecolor{YELLOW}{cmyk}{0,0,1,0}
}
\def \beq  {\begin{equation}}
\def \eeq  {\end{equation}}
\def \beqar {\begin{eqnarray}}
\def \eeqar {\end{eqnarray}}

\allowdisplaybreaks
\PassOptionsToPackage{caption=false}{subfig}
\newcommand{\Rmnum}[1]{\expandafter\@slowromancap\romannumeral #1@}

\makeatother

\usepackage{babel}
\begin{document}

\title{Effect of impurities on the Josephson current through helical metals: Exploiting a neutrino paradigm}

\author{Pouyan Ghaemi}
\affiliation{Physics Department, City College of the City University of New York, New York, NY 10031}

\author{V. P. Nair}
\affiliation{Physics Department, City College of the City University of New York, New York, NY 10031}

\begin{abstract}
In this letter we study the effect of time-reversal symmetric impurities on the Josephson supercurrent through  two dimensional helical metals such as on topological insulator surface state. We show that contrary to the usual superconducting-normal metal-superconducting junctions, the suppression of supercurrent in superconducting-helical metal-superconducting junction is mainly due to fluctuations of impurities in the junctions. Our results, which is a condensed matter realization of a part of the MSW effect for neutrinos, shows that the relationship between normal state conductance and critical current of Josephson junctions is significantly modified for Josephson junctions on the surface of topological insulators. We also study the temperature-dependence of supercurrent and present a two fluid model which can explain some of recent experimental results in 
Josephson junctions on the edge of topological insulators.
\end{abstract}

\maketitle

Critical currents in the mesoscopic superconducting-normal-superconducting (SNS) Josephson junctions have been studied widely\cite{JJreview} as an important property which can identify important physical properties of SNS Josephson junctions. Different regimes of coherent length, mean free path in the normal region and length of the normal region have been considered both experimentally and theoretically. One of the important results in this regard is the relationship between the normal state conductance and the critical current of the Josephson junctions \cite{kulik,beenakker,Bagwell,Foxon} which is well established both theoretically and experimentally. The discovery of  topological insulators (TIs) and the novel electronic band structure on their surface have led to many investigations on the unique features of electronic transport, such as quantum anti-localization, in this new class of materials\cite{RevModPhys.83.1057,RevModPhys.82.3045,NatPhys.5.378}. The experimental study of transport properties of the TI edge states 
has been quite challenging though as the bulk carriers contribute significantly to the transport in most of the experimentally accessible topological insulators. Superconductivity in TIs have been another widely studied area of research. The realization of superconductivity in doped TIs \cite{PhysRevLett.104.057001,PNAS.108.24}, as well as the possibility of making heterostructures of TIs with superconductors \cite{PhysRevLett.113.067003,GrowBiSE1} are important developments in this regard. Theoretical predictions such as topological superconductivity in doped TIs \cite{Fuberg} and presence of Majorana zero-mode\cite{NatPhys.5.614} in Josephson junctions through TI surface states \cite{FuMajorana,Zhao} motivated many experimental studies.  

The presence of helical edge states on the boundary of TIs is by now well established. The surface state band structure which resembles massless fermions, provides a unique platform to realize phenomena previously studied in high energy physics, such as axions\cite{Axion} and super-symmetry\cite{susy1,susy2}, in much more easily realizable condensed matter system.  
An important property of such states is the absence of back-scattering from electrostatic impurities,
which is enforced by the strong correlation between spin and momentum.
A reversal of momentum needs a reversal of spin and since electrostatic fields cannot flip
the spin, this helicity conservation forbids back-scattering. This, in turn, makes electronic transport through surface states of TIs insensitive to non-magnetic impurities.
Given this situation, it might seem that such impurities will not affect the supercurrent through 
surface states of TIs as well. 
In this letter we show that contrary to the situation for the normal state, non-magnetic impurities will affect the supercurrent carried by the surface states of TIs. This dynamical effect resembles yet another phenomenon familiar from high-energy physics in the context of neutrino oscillations known as Mikheyev-Smirnov-Wolfenstein (MSW) effect in topological insulators. We will show that the fluctuations of the impurities lead to a
a renormalization of the Fermi velocity.
This in turn means that the optical length of the junction (defined via the phase of the wave function)
is larger than the geometric length.
The modification of the phase also modifies (via matching conditions) the
energy eigenvalues of the Andreev states.  This is the essence of our result. In fact it has been noticed before that oscillations of impurities might affect the critical temperature in superconductors\cite{impmod}, but effect studied here as the sole mechanism for impurities to change the supercurrent in helical metals has not been considered before. Our results can be used to interpret the measurements on TI Josephson junctions which are currently the focus of many experimental studies \cite{ hor,vladimir,NatMat.11.417,PhysRevX.4.041022,MorpurgoBiSe,NadyaBiSe,PhysRevLett.109.056803}. 

The low-energy effective Hamiltonian of TI surface states reads as
\begin{equation}\label{sdh}
\mathcal{H}_s= v_F\,{\boldsymbol  \sigma} \cdot \textbf{k} 
\end{equation}
where  ${\boldsymbol  \sigma}=(\sigma_x,\sigma_y)$ are the Pauli matrices in the basis  $\left(\psi_\uparrow,\psi_\downarrow\right)$, with $\psi_\sigma$ being the electronic state with spin $\sigma$ localized on the surface of the TI. The low-energy effective Hamiltonian describing a Josephson junction on the surface of the TI, with supercurrent along $\hat{x}$, 
is given by \cite{pavan}
\begin{equation} \label{sh}
H=\left(-iv_F{\boldsymbol{\nabla}}\cdot\boldsymbol{\sigma}-\mu\right)\tau_3+\Delta_R(x)\tau_1+\Delta_I(x) \tau_2 ,
\end{equation} 
taken to act on the electronic state of the form
  $\left(\psi_\uparrow,\psi_\downarrow,\psi_\downarrow^\dagger,-\psi^\dagger_\uparrow\right)^T$. Here $\Delta_R(x)$ and $\Delta_I(x)$ are real and imaginary parts, respectively, of the induced superconducting gap $\Delta = \Delta_R +i \, \Delta_I$.
In (\ref{sdh}) and (\ref{sh}), $v_F$ is the Fermi velocity.
As for the matrix structure,  ${\sigma}_i$ act on physical spin space whereas the $\tau_i$ act on the superconducting particle-hole space. As the Hamiltonian in (\ref{sh}) is invariant under translation along $\hat{y}$, the momentum  $k_y$ in this direction is conserved. The low-energy Andreev states in the junction thus correspond to $k_y=0$ and $k_x$ close to the two fermi wave vectors,
 $k_x=\frac{\mu}{v_F}$ for $\sigma_x=1$ and $k_x=-\frac{\mu}{v_F}$ for $\sigma_x=-1$. Notice that since $\sigma_x$ commutes with Hamiltonian (\ref{sh}) we can decouple the low-energy effective Hamiltonians into two independent sectors corresponding to electron-hole states close to right or left fermi points. Here we will focus on one of the effective Hamiltonians, but the other independent one can be similarly studied. Since we are aiming for the effect of temporal fluctuations
 it is more efficient to use the corresponding $1+1$-dimensional action given by
\begin{equation}
S=\int dx dt \   \bar{\Phi}\left(x,t\right)\left[i \tau_1 D_t+ \tau_2  D_x +\tilde{M}\left(x\right)\right]\Phi\left(x,t\right)
\label{action1}
\end{equation}
where $\Phi^T\left(x,t\right)=\left(\phi_\uparrow\left(x,t \right),-\phi_\downarrow^\dagger\left(x,t\right)\right)$ and $\phi_\sigma\left(x,t\right)=\psi_\sigma\left(x,t\right) e^{\mp ik_F x}$, with $\sigma=\uparrow,\downarrow$, are the fermionic field operators for excitations close to the right or left Fermi points. $D_\mu=\partial_\mu-i\,e\,\tau_3 A_\mu$ is the covariant derivative.
Notice that (\ref{action1}) is of the standard form of a Dirac action
$\bar{\Phi} ( i \gamma^\mu D_\mu + {\tilde{M}} ) \Phi$ if we identify
$\gamma^0=\tau_1$, $\gamma^1=-i\tau_2$, $\gamma^5=\tau_3$.
Further, in (\ref{action1}), $\bar{\Phi}\left(x,t\right)=\Phi^\dagger\left(x,t\right)\gamma^0
= \Phi^\dagger (x,t) \, \tau_1$ and $\tilde{M}\left(x\right)=\Delta_R(x)\tau_0 + i \Delta_I(x)\tau_3$. 
The Fermi velocity has been set to $1$ by scaling $x$, or equivalently, the momentum $k_x$.
The effect of charged impurities is captured by $A_0=V(x-a(t))$ where $a(t)$ identifies instantaneous position of the impurity. As we will see below, in order to capture the effect of impurities on the
supercurrent in the junction, we should consider the natural fluctuations in the position of the impurity. For small fluctuations $a(t)=a_0+\xi(t)$, the impurity potential reads as $V(x-a(t))\approx V(x-a_0)+\partial_x V(x-a_0) \,\xi(t)$. As we will show below, the impurities can only affect the supercurrent as a result of  their temporal fluctuations. We would like to note that such treatment of impurities and its effect on superconductivity in normal metals have been considered long before \cite{impmod}. But here we show that as a result of helical band structure of the surface states of the TIs, the temporal fluctuations are the sole mechanism through which impurities can affect the supercurrent. The action, including the effect of fluctuating impurities, read as
\begin{eqnarray}
&&S=S_0+S_{int}+S_{osc}\nonumber\\
&&S_0=\int dx dt \Bigl[   \bar{\Phi}\left(x,t\right)\left( i \tau_1  \partial_t + \tau_2 \partial_x+\tilde{M}\left(x\right)\right) \Phi\left(x,t\right) \nonumber\\
&&\hskip .9in - V(x-a_0)\, \bar{\Phi}\left(x,t\right)i \tau_2 \Phi\left(x,t\right) \Bigr]\nonumber\\
&&S_{int}=\int dxdt \,\, \partial_x V(x-a_0)\,\, \bar{\Phi}\left(x,t\right)i\tau_2  {\Phi}\left(x,t\right) \,\xi(t)\nonumber\\
&&S_{osc}= {1\over 2} \,M_I \int dt\  \bar{\xi}(t) \left(-\frac{\partial^2}{\partial t^2}- \omega^2 \right) \xi(t)
\end{eqnarray}
Since we are interested in localized impurities, both the impurity potential $V(x-a)$ and the resulting electric field $E(x)=\partial_x V(x-a)$ are localized in space and the dynamics of the impurities are captured by the harmonic oscillator action $S_{osc}$ ($M_I$ and $\omega$ are the mass and the harmonic oscillation frequency of the impurities).  Coupling of Andreev states and the harmonic oscillations of impurities are captured in $S_{int}$ and will lead to a self-energy correction in $S_0$
(see supplementary materials), leading to an effective action of the form
\begin{equation}
\begin{split}
\bar{S}=& \int dx dt \  \Bigl[ \left( 1+ \Sigma_1(x) \right) \bar{\Phi} \left( i \tau_1 \partial_t + \tilde{M}\left(x\right)\right) \Phi  \\ 
&\hskip .6in + {\bar \Phi} \, \tau_2 \partial_x \Phi \, +\, {\bar\Phi}\, {\tilde M}(x) \Sigma_2 (x)
\,\Phi\Bigr] \\
&+\int dx dt\, V(x-a_0) \, \bar{\Phi}\, i\tau_2\, \Phi
\end{split}
\label{action2}
\end{equation}
To the lowest nontrivial order in perturbation theory, the self-energies can be calculated as
$\Sigma_1(x)\approx \frac{e^2 E^2(x)}{2\pi\omega^2 M_I}R$ and $\Sigma_2(x)\approx \frac{e^2 E^2(x)}{2\pi\omega^2 M_I}R\left[\log(\frac{2\omega}{M_I})-1\right]$ where $R$ is the length scale over which $E(x)= - \partial_x V(x-a)$ is non-zero (see supplementary materials). What is important for us is not so much the specific formulae for these self-energies, but that the general form of the effective action is as given
in (\ref{action2}), with the self-energies as corrections concentrated around the impurities.

Notice that the modification due to $\Sigma_1$ does not affect the spatial derivative term
for the electrons. This is because the oscillator variable $\xi (t)$ does not have
a spatial dependence. Thus although the free fermion action ${\bar \Phi} (i \gamma^\mu \partial_\mu
+ {\tilde M} ) \Phi$ has a Lorentz-type symmetry (albeit with $v_F$ in place of the speed of light $c$)
the interactions with impurities, and hence corrections, do not respect this symmetry.
The temporal and spatial derivative terms can be renormalized differently, and then with a
possible scaling of the field, the action can be brought to the form (\ref{action2}).
The general result is that the effect of fluctuating impurities appear as renormalization of 
the effective Fermi velocity and of the size of superconducting gap in the region where the electric field of the impurity potentials are present. The change of Fermi velocity can be viewed as a ``refractive index"
for the electron. There is then an additional phase change acquired for the wave functions 
as the effective ``optical length" is modified. This will be the essence of how the Andreev states are modified.

It is well known that for massless particles, propagating at the speed of light in vacuum,
the primary effect of interactions is to generate a refractive index rather than a mass
(which is usually forbidden for symmetry reasons). A similar situation is obtained even for massive particles in the ultrarelativistic limit. Our argument is that, for the surface states in a TI
which have a Lorentzian symmetry (with $c \rightarrow v_F$), again a refractive index is 
precisely what we should expect as the primary effect of interactions.

The situation here is closely related, conceptually, to how interactions with matter modify
neutrino oscillations as in the MSW effect \cite{Wolfenstein,Mikheyev1,Mikheyev2,Langacker}.
There are two ingredients to this. First, the neutrinos acquire a
refractive index which can be calculated in terms of the forward scattering amplitude,
a calculation which mirrors, {\it mutatis mutandis}, what is given in the supplementary material.
The refractive index for electron-neutrinos is different from that for other flavors
due to charged current interactions with matter.
The resulting difference in the phase of the wave functions modifies the neutrino mass-eigenstates
in matter, and hence the oscillations 
between different flavors of neutrinos.
The second part is a resonance effect which can enhance
the mixing of flavors in matter, even up to the maximal mixing. 
For us, we have only one flavor to consider, so the situation is simpler; there is no resonance part,
but the phase enhancement due to the refractive index is similar
and can  modify the
matching conditions (and energies) for the Andreev states.

 We will now use the effective action (\ref{action2})
to study the effect of impurities on the supercurrent. 
 The Andreev eigenstates are determined by the effective Schr\"odinger equation
which follows from (\ref{action2}),
\begin{equation}\begin{split}
- \left[\left(i\,\bar{v}\,\partial_x+  V(x-a_0)\right)\tau_3 +\Delta_R \tau_1+\Delta_I \tau_2\right] & \Psi(x)\\ = E\,&\Psi(x)
\end{split}\end{equation}
where  $\bar{v}(x)= {v_F/(1+{\Sigma_1(x)})}$ is the renormalized Fermi velocity. 
For constant $\Delta$, there are two independent eigenstates with energy $E$
given by
\begin{equation}\label{egs}
\begin{split}
\Psi^{\pm}_E(x)=&\,e^{i W^\pm(x) }\,\, \eta^\pm (E, \Delta )\\
W^\pm(x)=& \int_{0}^x du\, \,\frac{V(u-a_0)\pm \sqrt{E^2-|\Delta|^2}}{\bar{v}(u)}
\end{split}
\end{equation}
\begin{eqnarray}
 \eta^+(E,\Delta)&=&\frac{1}{\sqrt{2E}}\left[ \begin{array}{ccc} \sqrt{E+\sqrt{E^2-|\Delta|^2}} \\ -\Delta/\sqrt{E+\sqrt{E^2-|\Delta|^2}} \end{array} \right]\\
 \eta^-(E,\Delta)&=&\frac{1}{\sqrt{2E}}\left[ \begin{array}{ccc} -\Delta^*/\sqrt{E+\sqrt{E^2-|\Delta|^2}} \\ \sqrt{E+\sqrt{E^2-|\Delta|^2}} \end{array} \right]
\end{eqnarray}


To model the Josephson Junction, we consider the stepwise variation of $\Delta(x)$ in three regions $x<0$ (region I), $0<x<x_p$ (region II) and $x>x_p$ (region III) given by
\begin{equation}
\Delta(x)=\left\{\begin{array}{ccc }&\Delta_0&\  ({\rm I})\\ 
&0&\  ({\rm II}) \\ 
&\Delta_0  e^{i\phi_0} &\  ({\rm III}) \end{array} \right.
\end{equation}
The eigenstates can be expressed in each region as the superposition of $\Psi^{\pm}_E(x)$ as $\Psi_{i}(x)= A^i \Psi^+_E(x)+ B^i \Psi^-_E(x)$ where  $i= $ I, II and III, corresponding to the three regions.
We define the transfer matrices:
\begin{eqnarray}
\hat{T}_B (E,\Delta) &= &\frac{1}{\sqrt{2E}}
\left[ \begin{matrix}
\sqrt{E + \kappa v} & -\Delta^*/ \sqrt{E + \kappa v}\\
- \Delta/ \sqrt{E + \kappa v} & \sqrt{E + \kappa v} \\
 \end{matrix} \right]\nonumber\\
 \hat{T}_n(x)&= &\,e^{i\phi_I(x)}\left[\begin{array}{ccc} e^{i \langle k \rangle_x x} & 0 \\ 0 &  e^{-i  \langle k \rangle_x x}   \end{array}\right]\label{trnsfr}\\
 \hat{T}_s(x) &= & \left[\begin{array}{ccc} e^{i \kappa x} & 0 \\ 0 &  e^{-i\kappa x} \end{array}\right]
 \nonumber
\end{eqnarray}
where $\kappa= \sqrt{E^2-\Delta_0^2}\, /v$ is the wave vector in the superconducting region, $\langle k \rangle_x=\frac{E}{x} \int_0^x\frac{du}{\bar{v}(u)}$ is the averaged wave vector in the normal (TI) region
of the junction and $\phi_I(x)=\int_{0}^x \frac{V(u-a_0)}{\bar{v}(u)}  du$ is the phase resulting from the static impurity. In terms of these matrices, the eigenstates
in the three regions can be written as
\begin{eqnarray}
\Psi_{\rm I}(x)&=&\hat{T}_B(E,\Delta_0) \,\hat{T}_s(x) \, \left[\begin{array}{cc}  A^{\rm I} \\  
B^{\rm I} \end{array}\right]\\
\Psi_{\rm II}(x)&=&\hat{T}_n(x)\left[\begin{array}{cc}  A^{\rm II} \\  B^{\rm II} \end{array}\right]
\\
\Psi_{\rm III}(x)&=&\hat{T}_B(E,\Delta_0 e^{i\phi_0})\,  \hat{T}_s(x-x_p)\left[\begin{array}{cc}  A^{\rm III} \\  B^{\rm III} \end{array}\right]
\end{eqnarray}
The boundary conditions which determine the spectrum of the states in the
junction are
$\Psi_{\rm I}(0^-) = \Psi_{\rm II}(0^+)$ and  $\Psi_{\rm II}(x_p^-)=\Psi_{\rm III}(x_p^+)$.
These continuity equations reduce to
\begin{equation}\label{emain}
\left[\begin{array}{cc}  A^{\rm I} \\  B^{\rm I} \end{array}\right]=\hat{T}_J(E,\Delta,\phi_0) \left[\begin{array}{cc}  A^{\rm III} \\  B^{\rm III} \end{array}\right]
\end{equation}
where the wave function on the two ends of the Josephson junctions are connected by $\hat{T}_J(E,\Delta,\phi_0)=\hat{T}_B(E,\Delta_0)^{-1}\, \hat{T}_n(x_p)^{-1}\, \hat{T}_B(E,\Delta_0e^{i\phi_0})$. 
The $S$-matrix for the junction must relate the incoming and outgoing states as 
$\left[\begin{array}{cc}  A^{\rm III} \\  B^{\rm I} \end{array}\right]=\mathcal{S} \left[\begin{array}{cc}  A^{\rm I} \\  B^{\rm III} \end{array}\right]$; this $S$-matrix can be written in terms of  $\hat{T}_J(E,\Delta,\phi_0)$ as
\begin{equation}\label{smatrix}
\mathcal{S}= \frac{1}{\hat{T}_J(E,\Delta,\phi_0)_{11}}\left[\begin{array}{ccc} 1 & -\hat{T}_J(E,\Delta,\phi_0)_{12} \\ \hat{T}_J(E,\Delta,\phi_0)_{21}  &  e^{-2i\phi_I} \end{array}\right]
\end{equation}

The supercurrent in the junction can be derived using the well-known relationship between the Josephson current of the junction and the 
 spectrum \cite{beenakker}, namely,
\begin{eqnarray}
I&=& I_1+I_2+I_3\\
I_1&=&-\frac{e}{\hbar}\sum_p \tanh\left(E_p/2k_B T\right) \frac{d E_p}{d\phi} \label{I1}\\
I_2&=&-\frac{e}{\hbar}\int_{\Delta_0}^\infty \ dE\  \ln\left[2 \cosh\left(E/2k_BT\right)\right]\frac{\partial\rho(E,\phi)}{\partial \phi} \label{I2}\\
I_3&=&\frac{e}{\hbar}\frac{d}{d\phi} \int dx |\Delta(x)|^2/g\label{I3}
\end{eqnarray}
$I_1$ is the contribution form the discrete spectrum of in-gap states and $I_2$ is the contribution form continuum of states with energy above the gap with density of states $\rho(E,\phi)$ for the one spin state at each fermi point as we have for the helical metal. The third them $I_3$ vanishes for the phase independent gap and will be ignored in this letter. 
In (\ref{I3}), $g$ is the interaction constant of BCS theory.

 For the states with energy $E<\Delta_0$, $\kappa$ is imaginary and as a result $\Psi_E^+(x)$ in region I and $\Psi_E^-(x)$ in III are not normalizable. Equation (\ref{emain}) then simplifies to
\begin{equation}\label{emain2}
\left[\begin{array}{cc} 0\\  B^{\rm I} \end{array}\right]=\hat{T}_B(E,\Delta_0)^{-1}\hat{T}_n(x_p)^{-1}
\,\hat{T}_B(E,\Delta_0e^{i\phi_0})\left[\begin{array}{cc}  A^{\rm III} \\ 0 \end{array}\right]
\end{equation} 
and leads to the following equation determining the in-gap energies: 
\begin{equation}
\cos^{-1}(E_n/\Delta_0)+\frac{E_n}{v}
\mathcal{L} + { \phi_0\over 2} = n \,\pi, \hskip .2in n \in {\mathbb Z}
\label{egs2}
\end{equation}
where $\mathcal{L}=\int_0^{x_p}dx\,\left(1+\Sigma_1(x)\right)$ is the effective length of the junction which is modified by fluctuations of the impurity. This is the new ``optical length"
of the junction.
The phase factor $\phi_I$ has cancelled out in (\ref{egs2}) confirming that static impurities have no effect on the energy of in-gap Andreev states. The effect of impurities is only through their dynamical fluctuations that leads to the finite self-energy $\Sigma_1(x)$ which modifies the effective Fermi velocity. 
The effect on the energy eigenvalues is most vividly illustrated by considering states
with $E_n \ll \Delta_0$, in which case we get
$E_n \approx (v/ 2 {\mathcal L})  {\left[ 2\pi (n + {1\over 2} ) - \phi_0\right] }$.
The increase in ${\mathcal L}$ implies that $E_n$ and $\partial E_n /\partial \phi_0$
are decreased relative to the 
case with no impurities. 
More generally, defining $\Theta_n=\frac{E_n}{v}
\mathcal{L}+\phi_0/2-n\pi$, the supercurrent associated with each in-gap state reads as
\begin{equation}
{\mathcal I}_n(\phi_0)= - { e\over \hbar} {\partial E_n \over \partial \phi_0}
= \frac{e\,\vert \Delta_0 \vert}{2 \hbar}\left[\frac{\sin(\Theta_n)}{1+\sin(\Theta_n)\mathcal{L}|\Delta_0|/ v}\right]
\end{equation}
As a function of $\Theta_n$, this has a maximum at
$\Theta_n = \pi/2$, so that the critical current is 
${\mathcal I}_{crit} = (e \vert\Delta_0\vert /\hbar )(1+ {\mathcal L} \vert \Delta_0\vert /v)^{-1}$.
The condition $\Theta_n \approx \pi /2$ is actually obtained for modes of very low energy
$E_n \ll \vert \Delta_0\vert$. It is important to note that by that the suppercurrent generated by in gap states decreases by increasing $\mathcal{L}$ which shows that impurities clearly affect the supercurrent.



For the states above the gap, the density of states is given by the
Krein-Friedel-Lloyd formula $\rho(E)=\frac{1}{2\pi i} \frac{\partial}{\partial E} \left(\ln \det \mathcal{S} \right)$ \cite{kfl}. Using (\ref{smatrix}) we get 
\begin{equation}\nonumber
\det \mathcal{S}=e^{-i\phi_0}\frac{1- \beta_E^2\cos^2(\Theta_E)}{\left[ \sqrt{1-\beta_E^2}\,\cos(\Theta_E)- i\sin(\Theta_E)\right]^2}
\end{equation}
where $\Theta_E=\frac{E}{v} \mathcal{L}+\phi_0/2$ and $\beta_E=\frac{\Delta_0}{E}$. The
supercurrent due to above-the-gap states then simplifies to
\begin{eqnarray}
I_2 &= &  \frac{e}{4 \pi \,i\,\hbar} \left[\int_{\Delta_0}^\infty \ dE\   \tanh\left(E/2k_BT\right)\frac{\partial \left(\ln\det \mathcal{S}\right)}{ \partial \phi_0}\right]\nonumber\\
&&\hskip .2in  -\frac{e  k_B T}{2 \pi \hbar} \ln\left[2 \cosh\left(\Delta_0/2k_BT\right)\right] \nonumber\\
&= &  \frac{e}{4 \pi \hbar}\int_{\Delta_0}^\infty \!\!\! dE\   \tanh\left(E/2k_BT\right) \left[ \frac{E\sqrt{E^2-\Delta_0^2}}{E^2-\Delta_0^2 \cos^2(\Theta_E)}-1\right]\nonumber\\
&&\hskip .2in  -\frac{e  k_B T}{2 \pi \hbar} \ln\left[2 \cosh\left(\Delta_0/2k_BT\right)\right]
\label{absc}
\end{eqnarray}

We would like to emphasize two important features of the supercurrent contribution from states with energy above the superconducting gap:
\begin{enumerate}
\item For low temperatures $T\ll \Delta_0$, $I_2$ is only weakly $T$-dependent through the temperature dependence of superconducting gap $\Delta_0$.
\item $I_2$ is also only weakly dependent on $\mathcal{L}$, i.e, only weakly sensitive to
impurities.
\end{enumerate}

To elucidate the second point, we first note that the second term in (\ref{absc}) is independent of 
$\mathcal{L}$. Assuming $T\ll \Delta_0$, $\tanh\left(E/2k_BT\right) \sim 1$.  The integrand in the first term in (\ref{absc}) has two types of dependence on $E$. One is a periodic dependence, with period $\hbar v/\mathcal{L}$
due to $\cos(\Theta_E)$, and the other is a decaying dependence, of the form $\Delta_0^2/E^2$ for large $E$.
 For the effective junction length $\mathcal{L}$ larger than $\hbar v/\Delta_0$, the oscillatory dependence is is much faster than the decay rate and so can be averaged over $\Theta_E$. 
 (This may be viewed as an application of the Riemann-Lebesgue lemma.)
 As a result, the dependence on $\mathcal{L}$ will be eliminated and $I_2$ will not be seriously affected by impurities even when 
 fluctuation effects are included.

In conclusion we have shown that the supercurrent in Josephson junctions with helical metals, 
such as on the surface of three-dimensional TIs, is affected by
impurities through their temporal fluctuations. However, this applies primarily to
the supercurrent generated by in-gap Andreev states. The supercurrent carried by the states above the gap will not be seriously affected by impurities. Based on our results, the supercurrent in the Josephson junctions on the surface of TIs can be interpreted as a superposition of two contributions, one which is strongly temperature-dependent and also sensitive to the impurities in the junction and one which is only weakly temperature-dependent and 
not sensitive to the impurities. Given new advances in controlling the level of disorder in TIs
\cite{EviromentalDisorder1,EviromentalDisorder2}, these results will be useful in analyzing many of the experimental results on Josephson junctions made on TIs. For example, our analysis is consistent with the experimental results in \cite{vladimir,hor}; whether different level of impurities could affect the critical current in the Josephson junction on TI was the main missing ingredient in the theoretical model used to interpret those results.
In fact, our work may be considered as further substantiating the
interpretation, presented in \cite{hor},
 in terms of two types of supercurrent contributions. The detailed comparison with those experimental results will be subject of a following publication.

\bigskip
We would like to thank J. D. Sau, D. J. Van Harlingnen and A. Bernevig for helpful discussions. This work was supported in part by the U.S.\ National
Science Foundation grant PHY-1213380 and by a PSC-CUNY award.


\bibliography{TIJJ}

\begin{thebibliography}{35}%
\makeatletter
\providecommand \@ifxundefined [1]{%
 \@ifx{#1\undefined}
}%
\providecommand \@ifnum [1]{%
 \ifnum #1\expandafter \@firstoftwo
 \else \expandafter \@secondoftwo
 \fi
}%
\providecommand \@ifx [1]{%
 \ifx #1\expandafter \@firstoftwo
 \else \expandafter \@secondoftwo
 \fi
}%
\providecommand \natexlab [1]{#1}%
\providecommand \enquote  [1]{``#1''}%
\providecommand \bibnamefont  [1]{#1}%
\providecommand \bibfnamefont [1]{#1}%
\providecommand \citenamefont [1]{#1}%
\providecommand \href@noop [0]{\@secondoftwo}%
\providecommand \href [0]{\begingroup \@sanitize@url \@href}%
\providecommand \@href[1]{\@@startlink{#1}\@@href}%
\providecommand \@@href[1]{\endgroup#1\@@endlink}%
\providecommand \@sanitize@url [0]{\catcode `\\12\catcode `\$12\catcode
  `\&12\catcode `\#12\catcode `\^12\catcode `\_12\catcode `\%12\relax}%
\providecommand \@@startlink[1]{}%
\providecommand \@@endlink[0]{}%
\providecommand \url  [0]{\begingroup\@sanitize@url \@url }%
\providecommand \@url [1]{\endgroup\@href {#1}{\urlprefix }}%
\providecommand \urlprefix  [0]{URL }%
\providecommand \Eprint [0]{\href }%
\providecommand \doibase [0]{http://dx.doi.org/}%
\providecommand \selectlanguage [0]{\@gobble}%
\providecommand \bibinfo  [0]{\@secondoftwo}%
\providecommand \bibfield  [0]{\@secondoftwo}%
\providecommand \translation [1]{[#1]}%
\providecommand \BibitemOpen [0]{}%
\providecommand \bibitemStop [0]{}%
\providecommand \bibitemNoStop [0]{.\EOS\space}%
\providecommand \EOS [0]{\spacefactor3000\relax}%
\providecommand \BibitemShut  [1]{\csname bibitem#1\endcsname}%
\let\auto@bib@innerbib\@empty
\bibitem [{\citenamefont {Golubov}\ \emph {et~al.}(2004)\citenamefont
  {Golubov}, \citenamefont {Kupriyanov},\ and\ \citenamefont
  {Il'ichev}}]{JJreview}%
  \BibitemOpen
  \bibfield  {author} {\bibinfo {author} {\bibfnamefont {A.~A.}\ \bibnamefont
  {Golubov}}, \bibinfo {author} {\bibfnamefont {M.~Y.}\ \bibnamefont
  {Kupriyanov}}, \ and\ \bibinfo {author} {\bibfnamefont {E.}~\bibnamefont
  {Il'ichev}},\ }\href {\doibase 10.1103/RevModPhys.76.411} {\bibfield
  {journal} {\bibinfo  {journal} {Rev. Mod. Phys.}\ }\textbf {\bibinfo {volume}
  {76}},\ \bibinfo {pages} {411} (\bibinfo {year} {2004})}\BibitemShut
  {NoStop}%
\bibitem [{\citenamefont {Kulik}(1970)}]{kulik}%
  \BibitemOpen
  \bibfield  {author} {\bibinfo {author} {\bibfnamefont {I.~O.}\ \bibnamefont
  {Kulik}},\ }\href@noop {} {\bibfield  {journal} {\bibinfo  {journal} {Sov,
  Phys. JETP}\ }\textbf {\bibinfo {volume} {30}},\ \bibinfo {pages} {944}
  (\bibinfo {year} {1970})}\BibitemShut {NoStop}%
\bibitem [{\citenamefont {Beenakker}(1991)}]{beenakker}%
  \BibitemOpen
  \bibfield  {author} {\bibinfo {author} {\bibfnamefont {C.~W.~J.}\
  \bibnamefont {Beenakker}},\ }\href {\doibase 10.1103/PhysRevLett.67.3836}
  {\bibfield  {journal} {\bibinfo  {journal} {Phys. Rev. Lett.}\ }\textbf
  {\bibinfo {volume} {67}},\ \bibinfo {pages} {3836} (\bibinfo {year}
  {1991})}\BibitemShut {NoStop}%
\bibitem [{\citenamefont {Bagwell}(1992)}]{Bagwell}%
  \BibitemOpen
  \bibfield  {author} {\bibinfo {author} {\bibfnamefont {P.~F.}\ \bibnamefont
  {Bagwell}},\ }\href {\doibase 10.1103/PhysRevB.46.12573} {\bibfield
  {journal} {\bibinfo  {journal} {Phys. Rev. B}\ }\textbf {\bibinfo {volume}
  {46}},\ \bibinfo {pages} {12573} (\bibinfo {year} {1992})}\BibitemShut
  {NoStop}%
\bibitem [{\citenamefont {van Wees}\ \emph {et~al.}(1988)\citenamefont {van
  Wees}, \citenamefont {van Houten}, \citenamefont {Beenakker}, \citenamefont
  {Williamson}, \citenamefont {Kouwenhoven}, \citenamefont {van~der Marel},\
  and\ \citenamefont {Foxon}}]{Foxon}%
  \BibitemOpen
  \bibfield  {author} {\bibinfo {author} {\bibfnamefont {B.~J.}\ \bibnamefont
  {van Wees}}, \bibinfo {author} {\bibfnamefont {H.}~\bibnamefont {van
  Houten}}, \bibinfo {author} {\bibfnamefont {C.~W.~J.}\ \bibnamefont
  {Beenakker}}, \bibinfo {author} {\bibfnamefont {J.~G.}\ \bibnamefont
  {Williamson}}, \bibinfo {author} {\bibfnamefont {L.~P.}\ \bibnamefont
  {Kouwenhoven}}, \bibinfo {author} {\bibfnamefont {D.}~\bibnamefont {van~der
  Marel}}, \ and\ \bibinfo {author} {\bibfnamefont {C.~T.}\ \bibnamefont
  {Foxon}},\ }\href {\doibase 10.1103/PhysRevLett.60.848} {\bibfield  {journal}
  {\bibinfo  {journal} {Phys. Rev. Lett.}\ }\textbf {\bibinfo {volume} {60}},\
  \bibinfo {pages} {848} (\bibinfo {year} {1988})}\BibitemShut {NoStop}%
\bibitem [{\citenamefont {Qi}\ and\ \citenamefont
  {Zhang}(2011)}]{RevModPhys.83.1057}%
  \BibitemOpen
  \bibfield  {author} {\bibinfo {author} {\bibfnamefont {X.-L.}\ \bibnamefont
  {Qi}}\ and\ \bibinfo {author} {\bibfnamefont {S.-C.}\ \bibnamefont {Zhang}},\
  }\href@noop {} {\bibfield  {journal} {\bibinfo  {journal} {Rev. Mod. Phys.}\
  }\textbf {\bibinfo {volume} {83}},\ \bibinfo {pages} {1057} (\bibinfo {year}
  {2011})}\BibitemShut {NoStop}%
\bibitem [{\citenamefont {Hasan}\ and\ \citenamefont
  {Kane}(2010)}]{RevModPhys.82.3045}%
  \BibitemOpen
  \bibfield  {author} {\bibinfo {author} {\bibfnamefont {M.~Z.}\ \bibnamefont
  {Hasan}}\ and\ \bibinfo {author} {\bibfnamefont {C.~L.}\ \bibnamefont
  {Kane}},\ }\href@noop {} {\bibfield  {journal} {\bibinfo  {journal} {Rev.
  Mod. Phys.}\ }\textbf {\bibinfo {volume} {82}},\ \bibinfo {pages} {3045}
  (\bibinfo {year} {2010})}\BibitemShut {NoStop}%
\bibitem [{\citenamefont {Joel}(2009)}]{NatPhys.5.378}%
  \BibitemOpen
  \bibfield  {author} {\bibinfo {author} {\bibfnamefont {M.}~\bibnamefont
  {Joel}},\ }\href@noop {} {\bibfield  {journal} {\bibinfo  {journal} {Nature
  Physics}\ }\textbf {\bibinfo {volume} {5}},\ \bibinfo {pages} {378} (\bibinfo
  {year} {2009})}\BibitemShut {NoStop}%
\bibitem [{\citenamefont {Hor}\ \emph {et~al.}(2010)\citenamefont {Hor},
  \citenamefont {Williams}, \citenamefont {Checkelsky}, \citenamefont
  {Roushan}, \citenamefont {Seo}, \citenamefont {Xu}, \citenamefont
  {Zandbergen}, \citenamefont {Yazdani}, \citenamefont {Ong},\ and\
  \citenamefont {Cava}}]{PhysRevLett.104.057001}%
  \BibitemOpen
  \bibfield  {author} {\bibinfo {author} {\bibfnamefont {Y.~S.}\ \bibnamefont
  {Hor}}, \bibinfo {author} {\bibfnamefont {A.~J.}\ \bibnamefont {Williams}},
  \bibinfo {author} {\bibfnamefont {J.~G.}\ \bibnamefont {Checkelsky}},
  \bibinfo {author} {\bibfnamefont {P.}~\bibnamefont {Roushan}}, \bibinfo
  {author} {\bibfnamefont {J.}~\bibnamefont {Seo}}, \bibinfo {author}
  {\bibfnamefont {Q.}~\bibnamefont {Xu}}, \bibinfo {author} {\bibfnamefont
  {H.~W.}\ \bibnamefont {Zandbergen}}, \bibinfo {author} {\bibfnamefont
  {A.}~\bibnamefont {Yazdani}}, \bibinfo {author} {\bibfnamefont {N.~P.}\
  \bibnamefont {Ong}}, \ and\ \bibinfo {author} {\bibfnamefont {R.~J.}\
  \bibnamefont {Cava}},\ }\href@noop {} {\bibfield  {journal} {\bibinfo
  {journal} {Phys. Rev. Lett.}\ }\textbf {\bibinfo {volume} {104}},\ \bibinfo
  {pages} {057001} (\bibinfo {year} {2010})}\BibitemShut {NoStop}%
\bibitem [{\citenamefont {Zhanga}\ \emph {et~al.}(2010)\citenamefont {Zhanga}
  \emph {et~al.}}]{PNAS.108.24}%
  \BibitemOpen
  \bibfield  {author} {\bibinfo {author} {\bibfnamefont {J.~L.}\ \bibnamefont
  {Zhanga}} \emph {et~al.},\ }\href@noop {} {\bibfield  {journal} {\bibinfo
  {journal} {Proceedings of the National Academy of Sciences of the United
  States of America}\ }\textbf {\bibinfo {volume} {108}},\ \bibinfo {pages}
  {24} (\bibinfo {year} {2010})}\BibitemShut {NoStop}%
\bibitem [{\citenamefont {Yilmaz}\ \emph {et~al.}(2014)\citenamefont {Yilmaz},
  \citenamefont {Pletikosi\ifmmode~\acute{c}\else \'{c}\fi{}}, \citenamefont
  {Weber}, \citenamefont {Sadowski}, \citenamefont {Gu}, \citenamefont
  {Caruso}, \citenamefont {Sinkovic},\ and\ \citenamefont
  {Valla}}]{PhysRevLett.113.067003}%
  \BibitemOpen
  \bibfield  {author} {\bibinfo {author} {\bibfnamefont {T.}~\bibnamefont
  {Yilmaz}}, \bibinfo {author} {\bibfnamefont {I.}~\bibnamefont
  {Pletikosi\ifmmode~\acute{c}\else \'{c}\fi{}}}, \bibinfo {author}
  {\bibfnamefont {A.~P.}\ \bibnamefont {Weber}}, \bibinfo {author}
  {\bibfnamefont {J.~T.}\ \bibnamefont {Sadowski}}, \bibinfo {author}
  {\bibfnamefont {G.~D.}\ \bibnamefont {Gu}}, \bibinfo {author} {\bibfnamefont
  {A.~N.}\ \bibnamefont {Caruso}}, \bibinfo {author} {\bibfnamefont
  {B.}~\bibnamefont {Sinkovic}}, \ and\ \bibinfo {author} {\bibfnamefont
  {T.}~\bibnamefont {Valla}},\ }\href {\doibase 10.1103/PhysRevLett.113.067003}
  {\bibfield  {journal} {\bibinfo  {journal} {Phys. Rev. Lett.}\ }\textbf
  {\bibinfo {volume} {113}},\ \bibinfo {pages} {067003} (\bibinfo {year}
  {2014})}\BibitemShut {NoStop}%
\bibitem [{\citenamefont {Bansal}\ \emph {et~al.}(2011)\citenamefont {Bansal},
  \citenamefont {Kim}, \citenamefont {Edrey}, \citenamefont {Brahlek},
  \citenamefont {Horibe}, \citenamefont {Iida}, \citenamefont {Tanimura},
  \citenamefont {Li}, \citenamefont {Feng}, \citenamefont {Lee}, \citenamefont
  {Gustafsson}, \citenamefont {Andrei},\ and\ \citenamefont {Oh}}]{GrowBiSE1}%
  \BibitemOpen
  \bibfield  {author} {\bibinfo {author} {\bibfnamefont {N.}~\bibnamefont
  {Bansal}}, \bibinfo {author} {\bibfnamefont {Y.~S.}\ \bibnamefont {Kim}},
  \bibinfo {author} {\bibfnamefont {E.}~\bibnamefont {Edrey}}, \bibinfo
  {author} {\bibfnamefont {M.}~\bibnamefont {Brahlek}}, \bibinfo {author}
  {\bibfnamefont {Y.}~\bibnamefont {Horibe}}, \bibinfo {author} {\bibfnamefont
  {K.}~\bibnamefont {Iida}}, \bibinfo {author} {\bibfnamefont {M.}~\bibnamefont
  {Tanimura}}, \bibinfo {author} {\bibfnamefont {G.}~\bibnamefont {Li}},
  \bibinfo {author} {\bibfnamefont {T.}~\bibnamefont {Feng}}, \bibinfo {author}
  {\bibfnamefont {H.}~\bibnamefont {Lee}}, \bibinfo {author} {\bibfnamefont
  {T.}~\bibnamefont {Gustafsson}}, \bibinfo {author} {\bibfnamefont
  {E.}~\bibnamefont {Andrei}}, \ and\ \bibinfo {author} {\bibfnamefont
  {S.}~\bibnamefont {Oh}},\ }\href@noop {} {\bibfield  {journal} {\bibinfo
  {journal} {Thin Solid Films}\ }\textbf {\bibinfo {volume} {520}},\ \bibinfo
  {pages} {224} (\bibinfo {year} {2011})}\BibitemShut {NoStop}%
\bibitem [{\citenamefont {Fu}\ and\ \citenamefont {Berg}(2010)}]{Fuberg}%
  \BibitemOpen
  \bibfield  {author} {\bibinfo {author} {\bibfnamefont {L.}~\bibnamefont
  {Fu}}\ and\ \bibinfo {author} {\bibfnamefont {E.}~\bibnamefont {Berg}},\
  }\href {\doibase 10.1103/PhysRevLett.105.097001} {\bibfield  {journal}
  {\bibinfo  {journal} {Phys. Rev. Lett.}\ }\textbf {\bibinfo {volume} {105}},\
  \bibinfo {pages} {097001} (\bibinfo {year} {2010})}\BibitemShut {NoStop}%
\bibitem [{\citenamefont {Frank}(2009)}]{NatPhys.5.614}%
  \BibitemOpen
  \bibfield  {author} {\bibinfo {author} {\bibfnamefont {W.}~\bibnamefont
  {Frank}},\ }\href@noop {} {\bibfield  {journal} {\bibinfo  {journal} {Nature
  Physics}\ }\textbf {\bibinfo {volume} {5}},\ \bibinfo {pages} {614} (\bibinfo
  {year} {2009})}\BibitemShut {NoStop}%
\bibitem [{\citenamefont {Fu}\ and\ \citenamefont {Kane}(2008)}]{FuMajorana}%
  \BibitemOpen
  \bibfield  {author} {\bibinfo {author} {\bibfnamefont {L.}~\bibnamefont
  {Fu}}\ and\ \bibinfo {author} {\bibfnamefont {C.~L.}\ \bibnamefont {Kane}},\
  }\href@noop {} {\bibfield  {journal} {\bibinfo  {journal} {Phys. Rev. Lett.}\
  }\textbf {\bibinfo {volume} {100}},\ \bibinfo {pages} {096407} (\bibinfo
  {year} {2008})}\BibitemShut {NoStop}%
\bibitem [{\citenamefont {Olund}\ and\ \citenamefont {Zhao}(2012)}]{Zhao}%
  \BibitemOpen
  \bibfield  {author} {\bibinfo {author} {\bibfnamefont {C.~T.}\ \bibnamefont
  {Olund}}\ and\ \bibinfo {author} {\bibfnamefont {E.}~\bibnamefont {Zhao}},\
  }\href {\doibase 10.1103/PhysRevB.86.214515} {\bibfield  {journal} {\bibinfo
  {journal} {Phys. Rev. B}\ }\textbf {\bibinfo {volume} {86}},\ \bibinfo
  {pages} {214515} (\bibinfo {year} {2012})}\BibitemShut {NoStop}%
\bibitem [{\citenamefont {Qi}\ \emph {et~al.}(2008)\citenamefont {Qi},
  \citenamefont {Hughes},\ and\ \citenamefont {Zhang}}]{Axion}%
  \BibitemOpen
  \bibfield  {author} {\bibinfo {author} {\bibfnamefont {X.-L.}\ \bibnamefont
  {Qi}}, \bibinfo {author} {\bibfnamefont {T.~L.}\ \bibnamefont {Hughes}}, \
  and\ \bibinfo {author} {\bibfnamefont {S.-C.}\ \bibnamefont {Zhang}},\ }\href
  {\doibase 10.1103/PhysRevB.78.195424} {\bibfield  {journal} {\bibinfo
  {journal} {Phys. Rev. B}\ }\textbf {\bibinfo {volume} {78}},\ \bibinfo
  {pages} {195424} (\bibinfo {year} {2008})}\BibitemShut {NoStop}%
\bibitem [{\citenamefont {Ponte}\ and\ \citenamefont {Lee}(2014)}]{susy1}%
  \BibitemOpen
  \bibfield  {author} {\bibinfo {author} {\bibfnamefont {P.}~\bibnamefont
  {Ponte}}\ and\ \bibinfo {author} {\bibfnamefont {S.-S.}\ \bibnamefont
  {Lee}},\ }\href@noop {} {\bibfield  {journal} {\bibinfo  {journal} {New
  Journsl of Physics}\ }\textbf {\bibinfo {volume} {16}},\ \bibinfo {pages}
  {013044} (\bibinfo {year} {2014})}\BibitemShut {NoStop}%
\bibitem [{\citenamefont {Grover}\ \emph {et~al.}(2014)\citenamefont {Grover},
  \citenamefont {Sheng},\ and\ \citenamefont {Vishwanath}}]{susy2}%
  \BibitemOpen
  \bibfield  {author} {\bibinfo {author} {\bibfnamefont {T.}~\bibnamefont
  {Grover}}, \bibinfo {author} {\bibfnamefont {D.~N.}\ \bibnamefont {Sheng}}, \
  and\ \bibinfo {author} {\bibfnamefont {A.}~\bibnamefont {Vishwanath}},\
  }\href@noop {} {\bibfield  {journal} {\bibinfo  {journal} {Science}\ }\textbf
  {\bibinfo {volume} {344}},\ \bibinfo {pages} {280} (\bibinfo {year}
  {2014})}\BibitemShut {NoStop}%
\bibitem [{\citenamefont {Zhernov}\ \emph {et~al.}(1975)\citenamefont
  {Zhernov}, \citenamefont {Malov},\ and\ \citenamefont {Panova}}]{impmod}%
  \BibitemOpen
  \bibfield  {author} {\bibinfo {author} {\bibfnamefont {A.}~\bibnamefont
  {Zhernov}}, \bibinfo {author} {\bibfnamefont {Y.~A.}\ \bibnamefont {Malov}},
  \ and\ \bibinfo {author} {\bibfnamefont {G.~K.}\ \bibnamefont {Panova}},\
  }\href@noop {} {\bibfield  {journal} {\bibinfo  {journal} {Sov, Phys. JETP}\
  }\textbf {\bibinfo {volume} {42}},\ \bibinfo {pages} {131} (\bibinfo {year}
  {1975})}\BibitemShut {NoStop}%
\bibitem [{\citenamefont {Kurter}\ \emph {et~al.}(2014)\citenamefont {Kurter},
  \citenamefont {Finck}, \citenamefont {Ghaemi}, \citenamefont {Hor},\ and\
  \citenamefont {Van~Harlingen}}]{hor}%
  \BibitemOpen
  \bibfield  {author} {\bibinfo {author} {\bibfnamefont {C.}~\bibnamefont
  {Kurter}}, \bibinfo {author} {\bibfnamefont {A.~D.~K.}\ \bibnamefont
  {Finck}}, \bibinfo {author} {\bibfnamefont {P.}~\bibnamefont {Ghaemi}},
  \bibinfo {author} {\bibfnamefont {Y.~S.}\ \bibnamefont {Hor}}, \ and\
  \bibinfo {author} {\bibfnamefont {D.~J.}\ \bibnamefont {Van~Harlingen}},\
  }\href {\doibase 10.1103/PhysRevB.90.014501} {\bibfield  {journal} {\bibinfo
  {journal} {Phys. Rev. B}\ }\textbf {\bibinfo {volume} {90}},\ \bibinfo
  {pages} {014501} (\bibinfo {year} {2014})}\BibitemShut {NoStop}%
\bibitem [{\citenamefont {Orlyanchik}\ \emph {et~al.}()\citenamefont
  {Orlyanchik}, \citenamefont {Stehno}, \citenamefont {Nugroho}, \citenamefont
  {Ghaemi}, \citenamefont {Brahlek}, \citenamefont {Koirala}, \citenamefont
  {Oh},\ and\ \citenamefont {Harlingen}}]{vladimir}%
  \BibitemOpen
  \bibfield  {author} {\bibinfo {author} {\bibfnamefont {V.}~\bibnamefont
  {Orlyanchik}}, \bibinfo {author} {\bibfnamefont {M.~P.}\ \bibnamefont
  {Stehno}}, \bibinfo {author} {\bibfnamefont {C.~D.}\ \bibnamefont {Nugroho}},
  \bibinfo {author} {\bibfnamefont {P.}~\bibnamefont {Ghaemi}}, \bibinfo
  {author} {\bibfnamefont {M.}~\bibnamefont {Brahlek}}, \bibinfo {author}
  {\bibfnamefont {N.}~\bibnamefont {Koirala}}, \bibinfo {author} {\bibfnamefont
  {S.}~\bibnamefont {Oh}}, \ and\ \bibinfo {author} {\bibfnamefont {D.~J.~V.}\
  \bibnamefont {Harlingen}},\ }\href@noop {} {\enquote {\bibinfo {title}
  {Signature of a topological phase transition in the josephson supercurrent
  through a topological insulator},}\ }\bibinfo {note}
  {ArXiv:1309.0163}\BibitemShut {NoStop}%
\bibitem [{\citenamefont {Veldhorst}\ \emph {et~al.}(2012)\citenamefont
  {Veldhorst}, \citenamefont {Snelder}, \citenamefont {Hoek}, \citenamefont
  {Gang}, \citenamefont {Guduru}, \citenamefont {Wang}, \citenamefont
  {Zeitler}, \citenamefont {van~der Wiel}, \citenamefont {Golubov},
  \citenamefont {Hilgenkamp},\ and\ \citenamefont {Brinkman}}]{NatMat.11.417}%
  \BibitemOpen
  \bibfield  {author} {\bibinfo {author} {\bibfnamefont {M.}~\bibnamefont
  {Veldhorst}}, \bibinfo {author} {\bibfnamefont {M.}~\bibnamefont {Snelder}},
  \bibinfo {author} {\bibfnamefont {M.}~\bibnamefont {Hoek}}, \bibinfo {author}
  {\bibfnamefont {T.}~\bibnamefont {Gang}}, \bibinfo {author} {\bibfnamefont
  {V.~K.}\ \bibnamefont {Guduru}}, \bibinfo {author} {\bibfnamefont {X.~L.}\
  \bibnamefont {Wang}}, \bibinfo {author} {\bibfnamefont {U.}~\bibnamefont
  {Zeitler}}, \bibinfo {author} {\bibfnamefont {W.~G.}\ \bibnamefont {van~der
  Wiel}}, \bibinfo {author} {\bibfnamefont {A.~A.}\ \bibnamefont {Golubov}},
  \bibinfo {author} {\bibfnamefont {H.}~\bibnamefont {Hilgenkamp}}, \ and\
  \bibinfo {author} {\bibfnamefont {A.}~\bibnamefont {Brinkman}},\ }\href@noop
  {} {\bibfield  {journal} {\bibinfo  {journal} {Nature Materials}\ }\textbf
  {\bibinfo {volume} {11}},\ \bibinfo {pages} {417} (\bibinfo {year}
  {2012})}\BibitemShut {NoStop}%
\bibitem [{\citenamefont {Finck}\ \emph {et~al.}(2014)\citenamefont {Finck},
  \citenamefont {Kurter}, \citenamefont {Hor},\ and\ \citenamefont
  {Van~Harlingen}}]{PhysRevX.4.041022}%
  \BibitemOpen
  \bibfield  {author} {\bibinfo {author} {\bibfnamefont {A.~D.~K.}\
  \bibnamefont {Finck}}, \bibinfo {author} {\bibfnamefont {C.}~\bibnamefont
  {Kurter}}, \bibinfo {author} {\bibfnamefont {Y.~S.}\ \bibnamefont {Hor}}, \
  and\ \bibinfo {author} {\bibfnamefont {D.~J.}\ \bibnamefont
  {Van~Harlingen}},\ }\href {\doibase 10.1103/PhysRevX.4.041022} {\bibfield
  {journal} {\bibinfo  {journal} {Phys. Rev. X}\ }\textbf {\bibinfo {volume}
  {4}},\ \bibinfo {pages} {041022} (\bibinfo {year} {2014})}\BibitemShut
  {NoStop}%
\bibitem [{\citenamefont {Sacepe}\ \emph {et~al.}(2011)\citenamefont {Sacepe},
  \citenamefont {Oostinga}, \citenamefont {Li}, \citenamefont {Ubaldini},
  \citenamefont {Couto}, \citenamefont {Giannini},\ and\ \citenamefont
  {Morpurgo}}]{MorpurgoBiSe}%
  \BibitemOpen
  \bibfield  {author} {\bibinfo {author} {\bibfnamefont {B.}~\bibnamefont
  {Sacepe}}, \bibinfo {author} {\bibfnamefont {J.}~\bibnamefont {Oostinga}},
  \bibinfo {author} {\bibfnamefont {J.}~\bibnamefont {Li}}, \bibinfo {author}
  {\bibfnamefont {A.}~\bibnamefont {Ubaldini}}, \bibinfo {author}
  {\bibfnamefont {N.}~\bibnamefont {Couto}}, \bibinfo {author} {\bibfnamefont
  {E.}~\bibnamefont {Giannini}}, \ and\ \bibinfo {author} {\bibfnamefont
  {A.}~\bibnamefont {Morpurgo}},\ }\href@noop {} {\bibfield  {journal}
  {\bibinfo  {journal} {Nat. Commun.}\ }\textbf {\bibinfo {volume} {2}},\
  \bibinfo {pages} {575} (\bibinfo {year} {2011})}\BibitemShut {NoStop}%
\bibitem [{\citenamefont {Cho}\ \emph {et~al.}(2013)\citenamefont {Cho},
  \citenamefont {Dellabetta}, \citenamefont {Yang}, \citenamefont {Schneeloch},
  \citenamefont {Xu}, \citenamefont {Valla}, \citenamefont {Gu}, \citenamefont
  {Gilbert},\ and\ \citenamefont {Mason}}]{NadyaBiSe}%
  \BibitemOpen
  \bibfield  {author} {\bibinfo {author} {\bibfnamefont {S.}~\bibnamefont
  {Cho}}, \bibinfo {author} {\bibfnamefont {B.}~\bibnamefont {Dellabetta}},
  \bibinfo {author} {\bibfnamefont {A.}~\bibnamefont {Yang}}, \bibinfo {author}
  {\bibfnamefont {J.}~\bibnamefont {Schneeloch}}, \bibinfo {author}
  {\bibfnamefont {Z.}~\bibnamefont {Xu}}, \bibinfo {author} {\bibfnamefont
  {T.}~\bibnamefont {Valla}}, \bibinfo {author} {\bibfnamefont
  {G.}~\bibnamefont {Gu}}, \bibinfo {author} {\bibfnamefont {M.~J.}\
  \bibnamefont {Gilbert}}, \ and\ \bibinfo {author} {\bibfnamefont
  {N.}~\bibnamefont {Mason}},\ }\href@noop {} {\bibfield  {journal} {\bibinfo
  {journal} {Nat. Comm.}\ }\textbf {\bibinfo {volume} {4}},\ \bibinfo {pages}
  {1689} (\bibinfo {year} {2013})}\BibitemShut {NoStop}%
\bibitem [{\citenamefont {Williams}\ \emph {et~al.}(2012)\citenamefont
  {Williams}, \citenamefont {Bestwick}, \citenamefont {Gallagher},
  \citenamefont {Hong}, \citenamefont {Cui}, \citenamefont {Bleich},
  \citenamefont {Analytis}, \citenamefont {Fisher},\ and\ \citenamefont
  {Goldhaber-Gordon}}]{PhysRevLett.109.056803}%
  \BibitemOpen
  \bibfield  {author} {\bibinfo {author} {\bibfnamefont {J.~R.}\ \bibnamefont
  {Williams}}, \bibinfo {author} {\bibfnamefont {A.~J.}\ \bibnamefont
  {Bestwick}}, \bibinfo {author} {\bibfnamefont {P.}~\bibnamefont {Gallagher}},
  \bibinfo {author} {\bibfnamefont {S.~S.}\ \bibnamefont {Hong}}, \bibinfo
  {author} {\bibfnamefont {Y.}~\bibnamefont {Cui}}, \bibinfo {author}
  {\bibfnamefont {A.~S.}\ \bibnamefont {Bleich}}, \bibinfo {author}
  {\bibfnamefont {J.~G.}\ \bibnamefont {Analytis}}, \bibinfo {author}
  {\bibfnamefont {I.~R.}\ \bibnamefont {Fisher}}, \ and\ \bibinfo {author}
  {\bibfnamefont {D.}~\bibnamefont {Goldhaber-Gordon}},\ }\href@noop {}
  {\bibfield  {journal} {\bibinfo  {journal} {Phys. Rev. Lett.}\ }\textbf
  {\bibinfo {volume} {109}},\ \bibinfo {pages} {056803} (\bibinfo {year}
  {2012})}\BibitemShut {NoStop}%
\bibitem [{\citenamefont {Hosur}\ \emph {et~al.}(2011)\citenamefont {Hosur},
  \citenamefont {Ghaemi}, \citenamefont {Mong},\ and\ \citenamefont
  {Vishwanath}}]{pavan}%
  \BibitemOpen
  \bibfield  {author} {\bibinfo {author} {\bibfnamefont {P.}~\bibnamefont
  {Hosur}}, \bibinfo {author} {\bibfnamefont {P.}~\bibnamefont {Ghaemi}},
  \bibinfo {author} {\bibfnamefont {R.~S.~K.}\ \bibnamefont {Mong}}, \ and\
  \bibinfo {author} {\bibfnamefont {A.}~\bibnamefont {Vishwanath}},\ }\href
  {\doibase 10.1103/PhysRevLett.107.097001} {\bibfield  {journal} {\bibinfo
  {journal} {Phys. Rev. Lett.}\ }\textbf {\bibinfo {volume} {107}},\ \bibinfo
  {pages} {097001} (\bibinfo {year} {2011})}\BibitemShut {NoStop}%
\bibitem [{\citenamefont {Wolfenstein}(1978)}]{Wolfenstein}%
  \BibitemOpen
  \bibfield  {author} {\bibinfo {author} {\bibfnamefont {L.}~\bibnamefont
  {Wolfenstein}},\ }\href {\doibase 10.1103/PhysRevD.17.2369} {\bibfield
  {journal} {\bibinfo  {journal} {Phys. Rev. D}\ }\textbf {\bibinfo {volume}
  {17}},\ \bibinfo {pages} {2369} (\bibinfo {year} {1978})}\BibitemShut
  {NoStop}%
\bibitem [{\citenamefont {Mikheyev}\ and\ \citenamefont
  {Smirnov}(1985)}]{Mikheyev1}%
  \BibitemOpen
  \bibfield  {author} {\bibinfo {author} {\bibfnamefont {S.}~\bibnamefont
  {Mikheyev}}\ and\ \bibinfo {author} {\bibfnamefont {A.}~\bibnamefont
  {Smirnov}},\ }\href@noop {} {\bibfield  {journal} {\bibinfo  {journal} {Sov.
  J. Nucl. Phys.}\ }\textbf {\bibinfo {volume} {42}},\ \bibinfo {pages} {913}
  (\bibinfo {year} {1985})}\BibitemShut {NoStop}%
\bibitem [{\citenamefont {Mikheyev}\ and\ \citenamefont
  {Smirnov}(1986)}]{Mikheyev2}%
  \BibitemOpen
  \bibfield  {author} {\bibinfo {author} {\bibfnamefont {S.}~\bibnamefont
  {Mikheyev}}\ and\ \bibinfo {author} {\bibfnamefont {A.}~\bibnamefont
  {Smirnov}},\ }\href@noop {} {\bibfield  {journal} {\bibinfo  {journal} {Nuovo
  Cimento}\ }\textbf {\bibinfo {volume} {9C}},\ \bibinfo {pages} {17} (\bibinfo
  {year} {1986})}\BibitemShut {NoStop}%
\bibitem [{\citenamefont {Langacker}\ \emph {et~al.}(1983)\citenamefont
  {Langacker}, \citenamefont {Leveille},\ and\ \citenamefont
  {Sheiman}}]{Langacker}%
  \BibitemOpen
  \bibfield  {author} {\bibinfo {author} {\bibfnamefont {P.}~\bibnamefont
  {Langacker}}, \bibinfo {author} {\bibfnamefont {J.~P.}\ \bibnamefont
  {Leveille}}, \ and\ \bibinfo {author} {\bibfnamefont {J.}~\bibnamefont
  {Sheiman}},\ }\href {\doibase 10.1103/PhysRevD.27.1228} {\bibfield  {journal}
  {\bibinfo  {journal} {Phys. Rev. D}\ }\textbf {\bibinfo {volume} {27}},\
  \bibinfo {pages} {1228} (\bibinfo {year} {1983})}\BibitemShut {NoStop}%
\bibitem [{\citenamefont {Thirring}(1979)}]{kfl}%
  \BibitemOpen
  \bibfield  {author} {\bibinfo {author} {\bibfnamefont {W.}~\bibnamefont
  {Thirring}},\ }\href@noop {} {\emph {\bibinfo {title} {Quantum Mechanics of
  Atoms and Molecules: A Course in Mathematical Physics}}},\ Vol.~\bibinfo
  {volume} {3}\ (\bibinfo  {publisher} {Springer},\ \bibinfo {year}
  {1979})\BibitemShut {NoStop}%
\bibitem [{\citenamefont {Brahlek}\ \emph {et~al.}(2011)\citenamefont
  {Brahlek}, \citenamefont {Kim}, \citenamefont {Bansal}, \citenamefont
  {Edrey},\ and\ \citenamefont {Oh}}]{EviromentalDisorder1}%
  \BibitemOpen
  \bibfield  {author} {\bibinfo {author} {\bibfnamefont {M.}~\bibnamefont
  {Brahlek}}, \bibinfo {author} {\bibfnamefont {Y.~S.}\ \bibnamefont {Kim}},
  \bibinfo {author} {\bibfnamefont {N.}~\bibnamefont {Bansal}}, \bibinfo
  {author} {\bibfnamefont {E.}~\bibnamefont {Edrey}}, \ and\ \bibinfo {author}
  {\bibfnamefont {S.}~\bibnamefont {Oh}},\ }\href@noop {} {\bibfield  {journal}
  {\bibinfo  {journal} {Appl. Phys. Lett.}\ }\textbf {\bibinfo {volume} {99}},\
  \bibinfo {pages} {012109} (\bibinfo {year} {2011})}\BibitemShut {NoStop}%
\bibitem [{\citenamefont {Valdes}\ \emph {et~al.}(2013)\citenamefont {Valdes},
  \citenamefont {Wu}, \citenamefont {Stier}, \citenamefont {Bilbro},
  \citenamefont {Brahlek}, \citenamefont {Bansal}, \citenamefont {Oh},\ and\
  \citenamefont {Armitage}}]{EviromentalDisorder2}%
  \BibitemOpen
  \bibfield  {author} {\bibinfo {author} {\bibfnamefont {A.~R.}\ \bibnamefont
  {Valdes}}, \bibinfo {author} {\bibfnamefont {L.}~\bibnamefont {Wu}}, \bibinfo
  {author} {\bibfnamefont {A.~V.}\ \bibnamefont {Stier}}, \bibinfo {author}
  {\bibfnamefont {L.~S.}\ \bibnamefont {Bilbro}}, \bibinfo {author}
  {\bibfnamefont {M.}~\bibnamefont {Brahlek}}, \bibinfo {author} {\bibfnamefont
  {N.}~\bibnamefont {Bansal}}, \bibinfo {author} {\bibfnamefont
  {S.}~\bibnamefont {Oh}}, \ and\ \bibinfo {author} {\bibfnamefont {N.~P.}\
  \bibnamefont {Armitage}},\ }\href@noop {} {\bibfield  {journal} {\bibinfo
  {journal} {J. Appl. Phys.}\ }\textbf {\bibinfo {volume} {113}},\ \bibinfo
  {pages} {153702} (\bibinfo {year} {2013})}\BibitemShut {NoStop}%
\end{thebibliography}%

\pagebreak
\widetext

\setcounter{equation}{0}
\setcounter{figure}{0}
\setcounter{table}{0}
\setcounter{page}{1}


\def \beq  {\begin{equation}}
\def \eeq  {\end{equation}}
\def \beqar {\begin{eqnarray}}
\def \eeqar {\end{eqnarray}}
\def\ni{\noindent}
\def\be{\begin{equation}}
\def\ee{\end{equation}}
\def\bsp{\be\begin{split}}
\def\sqr#1#2{{\vcenter{\vbox{\hrule height.#2pt
\hbox{\vrule width.#2pt height#1pt \kern#1pt
\vrule width.#2pt}\hrule height.#2pt}}}}
\def\square{\mathchoice\sqr65\sqr65\sqr{5}3\sqr{5}3}

\def\adag{a^{\dagger}}
\def \bdag {b^{\dagger}}

\def\la {{\langle}}
\def\ra {{\rangle}}
\def\dag{\dagger}
\def\wt{\widetilde}
\def\wh{\widehat}

\def\vx {{\vec x}}
\def\vy {{\vec y}}
\def\vk {{\vec k}}
\def\vf {{\varphi}}
\def\bareta{{\bar \eta}}
\def\bvf {{\bar \varphi}}
\def\dag {{\dagger}}

\def\Tr {{\rm Tr}}
\def\tr{\mbox{tr}}

\def\tD{\widetilde{D}}
\def\tA{\widetilde{A}}
\def\tcD{{\widetilde{\cal D}}}

\def \dotA {{\dot A}}
\def \dotB {{\dot B}}

\def\bp {\bar p}
\def\ba {\bar{a}}
\def\bD {\bar{D}}
\def\bA {\bar{A}}
\def\bE {\bar{E}}
\def\bx {\bar{x}}
\def\by {\bar{y}}
\def\bz {{\bar{z}}}
\def\bu {\bar{u}}
\def\bv {\bar{v}}
\def\bw {\bar{w}}
\def\bV{{\bar V}}
\def\Jbar {{\bar J}}

\def\va {{\vec a}}
\def\vk {\vec{k}}
\def\vp {{\vec p}}
\def\vq {{\vec q}}
\def\vx {{\vec x}}
\def\vz {\vec{z}}
\def\vy{\vec{y}}
\def\vv {\vec{v}}
\def\vu {\vec{u}}
\def\vw {\vec{w}}

\def\vf {{\varphi}}
\def \bvf {{\bar \varphi}}
\def\btheta {\bar{\theta}}
\def\bxi {{\bar \xi}}
\def \blam {{\bar \lambda}}

\def\dag {\dagger}
\def\del {\partial}
\def\bdel{\bar{\partial}}

\def\a {\alpha}
\def\b {\beta}
\def\g{\gamma}
\def\d {\delta}
\def\e {\epsilon}
\def\k{\kappa}
\def\l {\lambda}
\def\m{\mu}
\def\n{\nu}
\def\r{\rho}
\def\s{\sigma}
\def\t{\tau}
\def\o {\omega}

\def\A {{\cal A}}
\def\C {{\cal C}}
\def\D {{\cal D}}
\def\E {{\cal E}}
\def\F {{\cal F}}
\def\G {{\cal G}}
\def\H {{\cal H}}
\def \L {{\cal L}}
\def\M{{\cal M}}
\def\O {{\cal O}}
\def\P {{\cal P}}
\def\S {{\cal S}}
\def\V {{\cal V}}

\def \lrdel {{\buildrel \leftrightarrow \over \partial}}
\def\half{\textstyle{1\over 2}}
\def\quarter {\textstyle{1\over 4}}
\def \onethird {{1\over 3}}
\def \twothird {{2\over 3}}
\def \hyph {{\penalty0\hskip0pt\relax}}

\def \tplus {{\tilde +}}
\def \tminus {{\tilde -}}

\def\theequation{S\arabic{equation}}



\begin{center}
\textbf{\large Supplementary Materials}
\end{center}

\section{Effective action for electrons from oscillating electrostatic impurity}
\subsection{The perturbative result}

Here we present the calculation of the effective action for the electron field due to the interaction with impurities.
The starting action is the one given in Eq.(3) in text. Since we will also be discussing how temperature affects the calculation, it is more convenient to use a Euclidean space (imaginary time) formalism, so that thermal effects can be included via the Matsubara formalism.

The Euclidean version of (3) is given by
\beq
S = \int d^2x ~\Phi^\dagger \, \left( D_4 - i \, \tau_3 D_1 \right) \Phi 
+ \int d^2x~ \Phi^\dagger \left[ \begin{matrix} 
0& \Delta \\  \Delta^*& 0\\ \end{matrix} \right] \Phi
\label{s1}
\eeq
Defining $\gamma^0 = \gamma^4 = \tau_1$, 
$\gamma^1 = - \tau_2$, this becomes
\beq
S = \int d^2x~ {\bar \Phi} \, \left[ \gamma^\mu ( \del_\mu - i \tau_3 \, A_\mu ) + M \right] \Phi
\label{s2}
\eeq
where ${\bar \Phi } = \Phi^\dagger \gamma^0$ and
\beq
M = \left[ \begin{matrix}
\Delta^* & 0\\ 0& \Delta \\ \end{matrix} \right]
\label{s3}
\eeq
The propagator for the electron is then given by
\beq
S(x,y) \equiv \la \Phi (x) \, {\bar \Phi}( y) \ra = \left( {1\over \gamma\cdot \del + M } \right)_{x,y}
= \int {d^2 p \over (2\pi )^2 } e^{-i p\cdot (x-y)} \, { ( i \gamma\cdot p + M^* )\over
( p^2 + M^* M)}
\label{s4}
\eeq
The electrostatic field can be expanded as
$A_4 (x + \xi )\approx A_4 (x) + \xi \del_1 A_4 
= A_4 (x) + \xi \, F_{14}$.
The correction to the action, to quadratic order is then given by the Wick contractions
of $- \half S_{int} S_{int} $; we need one electron propagator and one contraction for the $\xi$'s.
Thus
\beq
\Delta S_{eff} = e^2  \int  F_{14}(x) \, F_{14}(y)\,\la \xi (x) \xi (y) \ra \, {\bar \Phi} (x)
\gamma^0 \tau_3 \, S(x,y) \gamma^0 \tau_3 \Phi (y)
\label{s5}
\eeq
For us, $\xi$ depends only on time as it is the oscillating coordinate of the impurity relative to the
mean position. Further, if we consider several impurity atoms, only the $\xi$'s of the
same impurity atom can have nonzero average $\la \xi (x) \xi (y) \ra$. This means that
the contribution to the integral is concentrated around $x_1 = y_1$. We van encode these
by wriitng
\beq
\la \xi (x) \xi (y) \ra =  {R \over M_I} \int {d^2 k \over (2 \pi)^2} ~e^{-i k \cdot (x-y)}
{1\over k_4^2 + \omega^2}
\label{s6}
\eeq
The spatial part gives a delta function and integrating over $x_1$, we get the factor $R$.
This shows that $R$ may be taken as a rough measure of the extent over which this
correlation exists. Using this in (\ref{s5}) and changing variables $p = q -k$, we get
\beqar
\Delta S_{eff} &=& \int {d^2 q \over (2 \pi )^2} e^{-i q \cdot (x-y)} {\bar \Phi}(x) \,
\Sigma (q) \, \Phi (y)\label{s7}\\
\Sigma (q)&=& {e^2 R F_{14}^2 \over M_I} \int
{d^2 k \over (2 \pi)^2} { \gamma^0 \tau_3 \left( i \gamma\cdot ( q-k) +M^*\right) \gamma^0 \tau_3
\over (k_4^2 +\omega^2) [ (q-k)^2 + M^* M]}\nonumber\\
&=& {e^2 R F_{14}^2 \over M_I} \int_0^1 d \alpha~
\int{d^2 K \over (2 \pi)^2} {(i \gamma^0 q_4 - M ) (1-\alpha ) - \alpha M 
\over [ K_4^2 + \alpha K_1^2 + Q^2]^2}
\label{s8}\\
Q^2&=& \alpha (1 -\alpha ) q_4^2 + (1-\alpha ) \omega^2 + \alpha M^* M
\label{s9}
\eeqar
In the last line of (\ref{s8}) we have combined the denominators of the previous expression using
Feynman's formula and also defined $K_4 = k_4 - \alpha q_4$,
$K_1 = k_1 - q_1$. Notice that the self-energy $\Sigma$ is naturally split as
\beqar
\Sigma &=& - \Sigma_1 ( i \gamma^0 q_4 - M ) + \Sigma_2 \, M\nonumber\\
\Sigma_1 &=& -{e^2 R F_{14}^2 \over M_I} \int_0^1 d \alpha~
\int{d^2 K \over (2 \pi)^2} { (1-\alpha )
\over [ K_4^2 + \alpha K_1^2 + Q^2]^2}\label{s10}\\
\Sigma_2&=& - {e^2 R F_{14}^2 \over M_I} \int_0^1 d \alpha~
\int{d^2 K \over (2 \pi)^2} { \alpha 
\over [ K_4^2 + \alpha K_1^2 + Q^2]^2}\label{s11}
\eeqar
Evidently, the small $q$ limit of $\Sigma_1$ will give a term of the form ${\bar \Phi} (\gamma^0 \del_4 
+ M) \Phi$ in the action, while the same limit of $\Sigma_2$ will correct the mass $M$.
Higher order powers of $q_4$ in the expansion of $\Sigma$ will give higher derivatives of the electron
field; these are not important since we are interested in the low energy modes
of the electron. Explicitly,
\beq
\Delta S_{eff} \approx \Sigma_1 (q_4 = 0) \int {\bar \Phi} ( \gamma^0 \del_4 + M)  \, \Phi\, 
+ \Sigma_2 (q_4 =0) \int {\bar \Phi} M \Phi 
\label{s12}
\eeq
Carrying out the $K_4$ integration,
\beqar
\Sigma_1 (q_4 = 0 )&=& -{e^2 R F_{14}^2 \over M_I} \int_0^1 d \alpha~(1-\alpha )
\int{d K_1 \over 8 \pi} {1 \over [\alpha K_1^2 + \alpha M^*M + (1-\alpha ) \omega^2 ]^{3\over 2}}\nonumber\\
&\approx& {e^2  F_{10}^2  R\over M_I} {1\over 2 \pi \omega^2}\label{s13}
\eeqar
In the last line, we evaluated the remaining  integral neglecting $M^*M$ in comparison
to $\omega^2$, since the frequency of vibration for impurity atom is much larger than
the the possible gap. In any case, we set $M =)$ for the region II of our discussion in
text to which this calculation is applied.
We also made the continuation to real time by $F_{14}^2 \rightarrow F_{10}^2 = E^2$.
The result (\ref{s13}) agrees with what is quoted in text.

In a similar way,
\beq
\Sigma_2 (q_4 =0 ) \approx {e^2  E^2 R \over M_I} {1\over 2 \pi \omega^2}
\left[ \log (2 \omega / \vert \Delta\vert) - 1\right] \label{s14}
\eeq

We can include finite temperature effects by using Matsubara frequencies for
$k_4= 2 \pi n /\beta \equiv \omega_n$ and $p_4 = 2\pi (n+ \half) /\beta $,
$\beta = 1/(k_BT)$. The result is then
\beqar
\Delta S_{eff} &=& {1\over \beta}  \sum_s \int { d q_1 \over 2 \pi}  {\bar \Phi} e^{-i q \cdot (x-y)} 
\, \Sigma (q) \, \Phi (y) \nonumber\\
\Sigma(q) &=& {e^2 F_{14}^2 R \over M_I}
{1\over \beta} \sum_n \int {dk_1\over 2\pi} {i \gamma^0 ( q_{4s} - \omega_n ) + M^* + i \gamma^1 k_1
\over ( \omega_n^2 + \omega^2) [ ( q_{4s} - \omega_n )^2 + k_1^ 2 +M^*M]}
\label{s15}
\eeqar
As usual, we carry out the summation via contour integration. Writing
$Q^2 = k_1^2 + M^*M$,
\beq
{1\over \beta} \sum_n {i \gamma^0 ( q_{4s} - \omega_n ) + M^* + i \gamma^1 k_1
\over ( \omega_n^2 + \omega^2) [ ( q_{4s} - \omega_n )^2 + k_1^ 2 +M^*M]}
= {1\over 2 \pi} \oint {dz\over (e^{i \beta z} -1)}
{  i \gamma^0 (q_{4s} - z) + i \gamma^1 k_1 + M^* \over
(z^2 + \omega^2) [ (q_{4s}-z )^2 + Q^2]}
\label{s16}
\eeq
where the contour encloses the poles of the $(e^{i \beta z} - 1)^{-1}$ factor.
Folding the contours back to the upper and lower half-planes and evaluating the residues at the other poles, we find some terms which are independent of $\beta$ (and coincide with what we
have already done in (\ref{s8}- \ref{s14})) and a set of terms which are
$\beta$-dependent. The $\beta$-dependent part of $\Sigma_1$ is 
\beq
\Sigma_1\bigr]_{T\neq 0} =
{e^2 F_{14}^2 R \over M_I} \int {dk_1 \over 4 \pi \omega} 
\left[ {1\over Q^2 + (q_{4s}- i \omega)^2} + {1\over Q^2 + (q_{4s} + i \omega)^2}\right]
\, {1\over e^{\beta \omega} - 1}
\label{s17}
\eeq
The key point is that, for us, $\omega\gg k_B T$. As a result, the
$T$-dependent correction is exponentially suppressed  
due to the $e^{\beta \omega}$ factor in the denominator. 
A similar argument holds for $\Sigma_2$ as well.
Thus, in conclusion, the effective action is of the form as in Eq.(5) of the text where
we can take $\Sigma_1$ and $\Sigma_2$ to be independent of temperature.
Any temperature-dependence of the supercurrent would be due to the $T$-dependence of
the gap $\Delta$ and due to factors such as $\tanh (E/ 2 k_B T)$.

\subsection{The more general argument}

The form of the action in Eq.(5), which is all we need for the rest of the results in the paper, can be obtained on general symmetry grounds.
We will first consider
a region of uniform
distribution of impurities. 
The starting action has a $(1 +1)$-dimensional Lorentz symmetry with the
Fermi velocity in place of the speed of light.
If the interactions respect this symmetry, the relative coefficients
of the two terms  in the combination ${\bar\Phi} i \tau_1 \del_t \Phi
+ {\bar \Phi }\tau_2 \del_x \Phi$ must be preserved. However, the interactions do not preserve this
Lorentz symmetry and so, on general symmetry grounds, we expect the effective action to be of the form
\beqar
S_{eff} &=&  A  \int {\bar\Phi} i \tau_1 \del_t \Phi  + B \int {\bar \Phi }\tau_2 \del_x \Phi
+  \int {\bar \Phi } C \Phi \nonumber\\
&&\hskip .2in + {\rm terms ~ of ~ higher ~ order~in ~derivatives ~of}~\Phi
\label{s18}
\eeqar
The higher derivative terms are not important for the low energy modes which are of interest to us.
$A$, $B$, $C$ are calculable constants. 
We can scale out one of them, say, $B$ to write
\beq
S_{eff} =  (A/B)  \int {\bar\Phi} i \tau_1 \del_t \Phi  +  \int {\bar \Phi }\tau_2 \del_x \Phi
+  \int {\bar \Phi } (C/B) \Phi  + \cdots \label{s19}
\eeq
We now define $\Sigma_1$ 
\beq
A/B  = 1 + \Sigma_1
\label{s20}
\eeq 
With this, we see that the first term  (\ref{s19}) has the
form in Eq.(5). 

As for the ${\bar \Phi} {\tilde M} \Phi$ term, notice that the derivative terms of the starting action,
Eq. (3) in text, has a chiral symmetry, $\Phi \rightarrow e^{i \theta \tau_3} \Phi$.
This is true even with the electromagnetic interactions. Thus, if ${\tilde M} $ is originally
zero, it cannot be generated by perturbative corrections. Therefore, in the theory
with nonzero, $C$ must be such that it vanishes when ${\tilde M} \rightarrow 0$.
We therefore write $C \sim {\tilde M} $, and define $\Sigma_2$ in general by
\beq
{C\over B} = (1 +\Sigma_1 ) {\tilde M}  + {\tilde M}  \Sigma_2\label{s21}
\eeq
With the two definitions (\ref{s20}) and (\ref{s21}), we get
\beqar
S_{eff} &=&\int dx dt \  \Bigl[ \left( 1+ \Sigma_1(x) \right) \bar{\Phi} \left( i \tau_1 \partial_t + \tilde{M}\left(x\right)\right) \Phi  \nonumber\\
&&\hskip .6in + {\bar \Phi} \, \tau_2 \partial_x \Phi \, +\, {\bar\Phi}\, {\tilde M}(x) \Sigma_2 (x)
\,\Phi\Bigr]
\eeqar
in agreement with what is given in text.
The argument of the previous subsection was given to see how 
the perturbative calculation of the coefficients $A/B$ and $C/B$ 
can be done and give results consistent with the general expectations.

\end{document}